\documentclass[12pt]{iopart}
\usepackage{bm}
\usepackage{iopams}
\usepackage{setstack}
\usepackage{graphicx}

\begin{document}
\title[Analysis of OPM potentials for multiplet states 
of 3d transition metal atoms]
{Analysis of optimized effective potentials for multiplet states 
of 3d transition metal atoms}
\author{N Hamamoto and C Satoko}%
\address{%
Department of Integrated Sciences in Physics and Biology, 
College of Humanities and Sciences, Nihon University, 
3-25-40 Sakura-Jousui, Setagaya-ku, Tokyo 156-8550, Japan}
\ead{hamamoto@phys.chs.nihon-u.ac.jp}
\begin{abstract}
We apply the optimized effective potential method (OPM) to the 
multiplet energies of the 3d$^n$ transition metal atoms, 
where the orbital dependence of the energy functional 
with respect to orbital wave function is 
the single-configuration HF form.
%
We find that 
the calculated OPM exchange potential can be represented by the following 
two forms.
Firstly, the difference between OPM exchange potentials
of the multiplet states can be approximated by the 
linear combination of the potentials derived from the Slater integrals
$F^2({\rm 3d,3d})$ and $F^4({\rm 3d,3d})$ 
for the average energy of the configuration.
Secondly, the OPM exchange potential 
can be expressed as the linear combination
of the OPM exchange potentials of the single determinants.

\end{abstract}
\submitto{\jpb}
\maketitle
\section{Introduction}

The density functional theory (DFT) has a important problem with the
calculation of transition metal complexes. 
In spite of the recent development of the exchange-correlation
functionals, it is difficult to evaluate the multiplet energies using the
Kohn-Sham method,
because these electronic structures are not always written in a single
Slater determinant.
Several methods have been proposed to calculate the multiplet energies
using DFT.
One of these methods has been proposed by Ziegler \cite{Ziegler}, 
Wood \cite{Wood} and von Barth \cite{Barth}. 
They have carried out the computation of the multiplet energies
using the diagonal sum rule, in which
the sum of the multiplet energies is equal to 
the corresponding sum of the single determinant energies.
The method reproduces the multiplet energies of p electron system,
but the method does not give correct multiplet energies 
for the d electron system. \cite{Weinert}

The time dependent density functional theory (TDDFT) has been recently
developed to calculate excited states including the multiplet states.
The method has been applied to p electron systems \cite{TD-open}
within the adiabatic approximation.
The non-adiabatic TDDFT calculation, in which
the exchange-correlation kernel becomes  frequency dependent, is still
difficult.

Another method was formally discussed in the G\"{o}rling's paper
\cite{Gorl-sym} in which the energy functional depends on the
multiplet states.
They applied the optimized effective potential method 
to the calculation of the multiplet energies of p electron
atoms. \cite{Gorlingbase}
The optimized effective potential method (OPM) was firstly proposed by
Sharp and Horton\cite{Sharp}. The method was applied to the Neon and Carbon
atoms by Talman et. al.\cite{Talman}.
G\"{o}rling pointed out that OPM is formally regarded as DFT because
the orbital dependent energy functional implicitly depends on the density
\cite{densdiff}.
The OPM potential can be evaluated by the singular integral equation
called as the OPM equation.
Talman et. al.\cite{Talman,Aashamar} and Engel et. al.\cite{Engel}
solved the one-dimensional OPM equation for an atom, in which  
the angular components of the OPM equation for an atom are
integrated out using the spherical symmetry.
To efficiently calculate the OPM potentials for atoms and molecules, 
G\"{o}rling et. al. expanded the OPM potential using some base functions.
\cite{Gorlingbase,Ivanov,Ivanov2,Hirata}
The base expansion method gives the same total energy as
calculated by Talman's method.
However, in the base expansion method,
the OPM potential shows spurious oscillations.
To solve the problem, some methods have been proposed but
still are not applied to d electron atoms. \cite{Yang,Kummel,Kummel2}

An early work of OPM for the multiplet states is the OPM calculation
by Aashamar \cite{Aashamar} who applied OPM to 
the ground multiplet of the atom
of which atomic number is less than 85.
The OPM potentials for the p electron ground and excited state 
multiplets are calculated by Nagy \cite{Nagy} 
using the KLI method.\cite{KLI,KLImol}
Furthermore, Aashamar also applied OPM to the multiplet energies of 
p electron atoms using the multi-configuration HF form. \cite{AaMCSCF} 
The total energies of the multiplet states have been
discussed in these studies, but
the potentials of the multiplet states are not presented.
Sala and G\"{o}rling have been developed the LHF method to approximately
calculate the exchange only OPM potential. \cite{LHF,LHFasym1,LHFasym2}
Recently, the LHF method is extended to 
the open shell atoms and molecules.\cite{LHFopen}
They applied the LHF method to the ground and excited 
multiplets of p electron atoms and molecules.
However, in these papers, they did not treat the excited multiplet energies of 
the d electron atoms, which are difficult to calculate using the DFT.
Furthermore, these studies give incorrect result
caused by the approximation methods such as 
 the KLI or the LHF method.

In the present paper, we apply OPM to the ground and excited state multiplets
of the 3d$^n$ electron atoms using the total energy
functional of the single-configuration HF  \cite{Fischer}  form.
Following Talman and Engel\cite{Talman,Engel}, we solve the 
one-dimensional OPM equation correctly. 
The solution is considered to give the most reliable 
result among all the available methods.
We show that the differences between the OPM potentials of 
the multiplet states can be approximated by the linear combination 
of the potentials derived from $F^2({\rm 3d,3d})$,$F^4({\rm 3d,3d})$ Slater integrals.
We numerically show that the OPM exchange potential of multiplet state is
approximated by the linear combination of the OPM exchange
potentials of the single determinants.

\section{Optimized Effective Potential Method}

In the present paper, we calculate the OPM potential for eigenstates
of an transition metal atom. 
Each state is characterized by a
definite value of the total orbital angular momentum $L$ and of 
the total spin angular momentum $S$.
These multiplet energies can be expressed as  $E^{LS,s}$, where
$s$ is the seniority number.
The multiplet energy level is degenerate for possible directions of
total momentums $L$ and $S$.
Then, each multiplet wavefunction $\Psi(LSL_z S_z,s)$ can be 
written as a linear combination of single Slater determinant
functions $\Phi(L_z S_z,\nu)$ which are not always 
eigenfunctions of the atom.
$L_z$ and $S_z$ are $z$ component of the total momentums $L$ and $S$,
respectively, and $\nu$ specifies different states
with the same $L_z$ and $S_z$ component.
Each total energy of the single determinant is represented as
 $E^{L_z S_z,\nu}$.

Furthermore, we consider ``average energy of the configuration'' $E^{\rm av}$ 
defined by Slater. \cite{Slater,Fischer} 
For the 3d$^n$ multiplets, $E^{\rm av}$ is expressed as
\begin{equation}
E^{\rm av} = \frac{(10-n)! n!}{10!}\sum_{LS,s}(2L+1)(2S+1)E^{LS,s}.
\label{average_energy}
\end{equation}

To generally express these three kind of energies,
 $E^{LS,s}$,$E^{L_z S_z,\nu}$ and $E^{\rm av}$, 
we introduce $E^\xi$ where $\xi$ stands for
$\{LS,s\}$,$\{L_z S_z, \nu\}$ and $\{{\rm av}\}$.
The total energy $E^\xi$ is divided into the 1-body part  $E_1^\xi$ 
and the 2-body part $E_2^\xi$;
\begin{equation}
\label{total_energy}
E^\xi = E^\xi_1 + E_2^\xi.
\end{equation}
The 1-body part is constructed from the kinetic energy of 
electrons and the electron-nucleus Coulomb energy.
\begin{eqnarray}
\label{eng_1body}
E_1^\xi = \sum_i q_i I^\xi(n_i l_i),\\
\label{eng_1body_sub}
I^\xi(n_i l_i) = 
\int P^\xi_{n_i l_i}(r) {\cal L}_i P^\xi_{n_i l_i}(r) dr,\\
{\cal L}_i = ( - \frac{1}{2}\frac{d^2}{dr^2} 
                      + \frac{l_i(l_i +1)}{2 r^2}
                      - \frac{Z}{r} ),
\end{eqnarray}
where $q_i$ represents the number of the electrons in 
the orbital $P^\xi_{n_i l_i}(r)$.
$Z$ is the atomic number.
$n_i$ and $l_i$ are the principal quantum number and
 the angular quantum number, respectively.
In this study, we suppose that the OPM potential $V^{\xi}_{\rm OPM}(r)$
is spherically symmetric.
The single electron orbital is determined by the following equation
\begin{eqnarray}
\label{P-equation}
({\cal L}_i + V_{\rm OPM}^{\xi}(r)) P^\xi_{n_il_i}(r) = 
\epsilon^\xi_i  P^\xi_{n_il_i}(r),
\end{eqnarray}
where $\epsilon^\xi_i$ is the single electron energy.
$P^\xi_{n_i l_i}(r)$ is the product of the distance from 
the nucleus $r$ and the radial component of the single electron wave function.

The 2-body part $E_2^\xi$ is composed of the electron-electron interaction
energy expressed as
\begin{eqnarray}
\label{eng_2body}
E_2^\xi = \frac{1}{2} \sum_{ijk} a_{ijk}^\xi F^k_\xi(ij)
+\frac{1}{2} \sum_{i\neq j, k} b_{ijk}^\xi G^k_\xi(ij),
\end{eqnarray}
where the coefficients $a_{ijk}^\xi$ and $b_{ijk}^\xi$ depend on
the multiplet states, the single determinants or the average energy of the
configuration.
The Slater integrals $F_\xi^k(ij)$ and $G_\xi^k(ij)$ are defined as
\begin{eqnarray}
F^k_\xi(ij) &=& \int dr \int dr'
P^\xi_{n_il_i}(r)P^\xi_{n_il_i}(r) \frac{r_<^k}{r_>^{k+1}}
P^\xi_{n_jl_j}(r')P^\xi_{n_jl_j}(r'),\\
G^k_\xi(ij) &=& \int dr \int dr' P^\xi_{n_il_i}(r)P^\xi_{n_jl_j}(r)
\frac{r_<^k}{r_>^{k+1}} P^\xi_{n_jl_j}(r')P^\xi_{n_il_i}(r'),
\end{eqnarray}
where $r_<$ is the smaller of $r$ , $r'$ and $r_>$ is the larger.

The OPM potential $V_{\rm OPM}^\xi(r)$ is determined by requiring
that $E^\xi$ be minimized for all $P^\xi_{n_i l_i}$ obtained from
equation (\ref{P-equation}). This results in
\begin{eqnarray}
\label{OEPeq-org1}
\frac{\delta E^\xi}{\delta V_{\rm OPM}^\xi(r)} 
     = \sum_i \int dr'
\frac{\delta E^\xi} {\delta P^\xi_{n_i l_i}(r')} 
\frac{\delta P^\xi_{n_i l_i}(r')}{\delta V_{\rm OPM}^\xi(r)} = 0,
\end{eqnarray}
where
\begin{equation}
\label{OEPeq-org2}
\frac{\delta E^\xi}{\delta P^\xi_{n_i l_i}(r')} =
2 q_i {\cal L}_i P^\xi_{n_i l_i}(r')+
\frac{\delta E_2^\xi}{\delta P^\xi_{n_i l_i}(r')}.
\end{equation}
Using equations (\ref{total_energy}),(\ref{P-equation}),(\ref{OEPeq-org2}),
and the variation of the normalization condition 
$\int P^\xi_{n_i l_i}(r)^2 dr=1$ with respect to 
$V_{\rm OPM}^\xi (r')$, that is, 
$\int P^\xi_{n_i l_i}(r)\frac{\delta P^\xi_{n_i l_i}(r)}
{\delta V_{\rm OPM}^\xi (r')}dr =0$, 
we can rearrange equation (\ref{OEPeq-org1}) as
\begin{eqnarray}
\label{OEPeq-org3}
\sum_i \int dr'
\left( 
2 q_i V_{\rm OPM}^\xi(r) P^\xi_{n_i l_i}(r')-
\frac{\delta E_2^\xi}{\delta P^\xi_{n_i l_i}(r')}
\right)
\frac{\delta P^\xi_{n_i l_i}(r')}{\delta V^\xi_{\rm OPM}(r)} = 0.
\end{eqnarray}
The functional derivative
$\frac{\delta P^\xi_{n_i l_i}(r')}{\delta V^\xi_{\rm OPM}(r)}$
appeared in equation (\ref{OEPeq-org1}) is calculated using the perturbation
theory:
\begin{eqnarray}
\label{OEPeq-org4}
\frac{\delta P_{n_i l_i}^\xi(r')}{\delta V_{\rm OPM}^\xi(r)}
= -G_i^\xi(r,r')P^\xi_{n_i l_i}(r),
\quad G^\xi_i(r,r')=\sum_{j \neq i} 
\frac{P^\xi_{n_j l_j}(r')P^\xi_{n_j l_j}(r)}
{\epsilon^\xi_j -\epsilon^\xi_i}.
\end{eqnarray}
The functional derivative
$\frac{\delta E_2^\xi}{\delta P^\xi_{n_i l_i}(r')}$ 
in equation (\ref{OEPeq-org3}) is calculated as
\begin{eqnarray}
\label{OEPeq-org5}
\frac{\delta E_2^\xi}{\delta P^\xi_{n_i l_i}(r')}
=\sum_{jk} 2 a_{ijk}^{\xi} X^\xi_k(jj, r') P^\xi_{n_i l_i}(r')
+\sum_{jk} 2 b_{ijk}^{\xi} X^\xi_k(ij, r') P^\xi_{n_j l_j}(r'),
\end{eqnarray}
where $X^\xi_k(ij, r)$ is defined as
$X^\xi_k(ij, r) = \int P^\xi_{n_il_i}(r') P^\xi_{n_jl_j}(r') 
\frac{r_<^{k}}{r_>^{k+1}} dr'$.
Substituting equations (\ref{OEPeq-org4}) and (\ref{OEPeq-org5}) 
into equation (\ref{OEPeq-org3}), we obtain
\begin{eqnarray}
\sum_i 
\int dr'
\left(
q_i V_{\rm OPM}^\xi(r')
- \sum_{jk} a_{ijk}^\xi X^\xi_k(jj,r')
\right) 
P^\xi_{n_i l_i}(r') G^\xi_i(r,r')P^\xi_{n_i l_i}(r) \nonumber\\
=\sum_{i}\sum_{jk} \int b^\xi_{ijk} X^\xi_k(ij,r') 
P^\xi_{n_j,l_j}(r') G^\xi_i(r,r')P^\xi_{n_i l_i}(r) dr'.
\label{OEP-equation}
\end{eqnarray}
Equation (\ref{OEP-equation}) is called as the OPM equation. 
The OPM potential can be obtained by self-consistently solving
the OPM equation and the single electron equation (\ref{P-equation}).

\section{Method of Calculations}
\label{cal-method}

To solve the OPM equation (\ref{OEP-equation}), we modified the
code developed by Fischer et. al. \cite{Fischer}.
For the calculation of the Green's function, we use the following expression
\begin{eqnarray}
G^\xi_i(r,r')&=&P^\xi_{n_i l_i}(r_>)Q^\xi_{n_i l_i}(r_<)
-P^\xi_{n_i l_i}(r')\Phi^\xi_{n_i l_i}(r)-P^\xi_{n_i l_i}(r)\Phi^\xi_{n_i l_i}(r')\nonumber\\
&&+C^\xi_{n_i l_i}P^\xi_{n_i l_i}(r)P^\xi_{n_i l_i}(r'),
\end{eqnarray}
where $Q^\xi_{n_i l_i}(r)$ is the second solution of equation (\ref{P-equation})
satisfying $\frac{d P^\xi_{n_i l_i}(r)}{dr}Q^\xi_{n_i l_i}(r)
-\frac{d Q^\xi_{n_i l_i}(r)}{dr}P^\xi_{n_i l_i}(r)=1$, and
\begin{eqnarray}
\Phi^\xi_{n_i l_i}(r) &=& 
  P^\xi_{n_i l_i}(r) \int_0^r      P^\xi_{n_i l_i}(r')Q^\xi_{n_i l_i}(r')dr'
\nonumber\\
&&+ Q^\xi_{n_i l_i}(r) \int_r^\infty P^\xi_{n_i l_i}(r')P^\xi_{n_i l_i}(r')dr',
\end{eqnarray}
with $C^\xi_{n_i l_i}=\int P^\xi_{n_i l_i}(r)\Phi^\xi_{n_i l_i}(r)dr$.
To evaluate $P^\xi_{n_i l_i}(r)$ and $Q^\xi_{n_i l_i}(r)$,
we solve the single electron equation (\ref{P-equation}) 
using the Numerov's method with the transformation
$h=\log(Zr)$ and $\bar{P}_{n_i l_i}(h)=P_{n_i l_i}(r)/\sqrt{r}$.
The lower bound of $h$ is set to -6.0 and the upper bound, which 
depends on the kind of atom, is set around 7. The step size
$\Delta h$ is $0.005$.
The integration in the OPM equation is approximated 
as the sum of the product of integrand and step size.
Using this simple approximation with the step size $\Delta h=0.01$,
the OPM equation is approximated to the set of linear equations for the 
OPM potential.
The OPM equation (\ref{OEP-equation}) determines the solution
$V_{\rm OPM}^\xi(r)$ only up to a constant. The constant is fixed by
the physical requirement 
$\lim_{r \rightarrow \infty} V_{\rm OPM}^\xi(r)=0$.
In our program, we fix the constant using the asymptotic form of 
the OPM potential
$-\sum_{k}\int (a_{NNk}^\xi + b_{NNk}^\xi)
\frac{r_<^k}{r_>^{k+1}}{P^\xi_{n_N l_N}}^2(r') dr'$ at $r=9$ au,
where $N$ is the index for HOMO.
The self-consistent procedure converges satisfactorily in about 12
iterations if the average of the initial and final $V_{\rm OPM}(r)$ is
taken at each iteration. The SCF iteration has converged when 
$\max_i (\sqrt{q_i}\Delta P_{n_i l_i})
< 1.0\times 10^{-8}\sqrt{Z N_{\rm occ}}$, 
where $N_{\rm occ}$ is the number of the occupied orbitals. \cite{Fischer}

In OPM, the single electron energy of HOMO $\epsilon_{\rm OPM}^\xi$ 
is equal to the HF single electron expectation value for HOMO 
$\epsilon_{\rm HF}^\xi$\cite{KLI}.
In our results of the calculations for the 3d transition metal atoms, 
the differences between the single particle energies
($|\epsilon_{\rm OPM}^\xi-\epsilon_{\rm HF}^\xi|$) are
less than 0.005 au.

\section{Results and Discussions}

The OPM exchange potential $V^\xi_{\rm ex}(r)$ 
is defined by subtracting spherical average of the Coulomb potential
from the OPM potential $V^\xi_{\rm OPM}(r)$:
\begin{eqnarray}
V^\xi_{\rm ex}(r)=V^\xi_{\rm OPM}(r)-\int \frac{\rho^\xi(r')}{r_>}dr',
\end{eqnarray}
where the radial density $\rho^\xi(r)$ is defined as
\begin{equation}
\label{density-def}
\rho^\xi(r)=\sum_i q_{n_i} P^\xi_{n_i l_i}(r)^2.
\end{equation}

Firstly, we apply OPM to the average energy $E^{\rm av}$ defined by
equation (\ref{average_energy}).
The average OPM potential $V^{\rm av}_{\rm OPM}(r)$ and average single
electron orbital $P^{\rm av}_{n_il_i}(r)$ are obtained by the
self-consistent solution of the OPM equation (\ref{OEP-equation}) and
the single electron equation (\ref{P-equation}).  The average radial density
$\rho^{\rm av}(r)$ is calculated using equation (\ref{density-def}).
In DFT, the exchange energy is expressed as
$E_{\rm DFTex}[n(\bm{r})]$ where $n(\bm{r})$ is an electron density.
To calculate averaged DFT potential, we used spherical averaged density 
 $\rho^{\rm av}(r)/4\pi r^2$ denoted as $n^{\rm av}(\bm{r})$.
Then, the average energy for DFT is defined as
\begin{eqnarray}
\label{av-Eng-DFT}
E^{\rm av}_{\rm DFT}=\sum_i q_i I^{\rm av}(n_i l_i) 
+ \int \frac{n^{\rm av}(\bm{r}) n^{\rm av}(\bm{r'})}
{|\bm{r}-\bm{r'}|}d\bm{r}d\bm{r'}
+ E_{\rm DFTex}[n^{\rm av}(\bm{r})].
\end{eqnarray}
The average density is calculated by the self-consistent solution
of the Kohn-Sham equation with spherical exchange potential 
$\frac{\delta E_{\rm DFTex}[n^{\rm av}(\bm{r})]}{\delta n^{\rm av}(\bm{r})}$.

\begin{figure}
\begin{center}
\includegraphics[angle=-90,scale=.5]{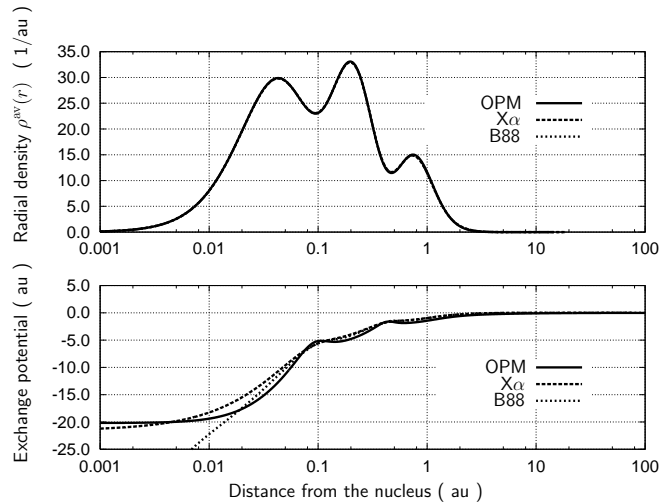}
\caption{
\label{DFTpot_Mn}
The radial density $\rho^{\rm av}(r)$ of Mn$^{2+}$ 
calculated using the X$\alpha$, B88 and OPM exchange potentials
 for the average energy (top).
The X$\alpha$, B88 and OPM exchange potentials
for the average energy of Mn$^{2+}$ (bottom).
}
\end{center}
\end{figure}

In the bottom part of figure \ref{DFTpot_Mn}, we show 
the X$\alpha$ exchange potential ($\alpha=2/3$), 
the Becke's GGA (B88) exchange potential \cite{Becke88},
and the OPM exchange potential $V_{\rm ex}^{\rm av}(r)$ of Mn$^{2+}$.
The OPM exchange potential is close to the X$\alpha$ and B88 
exchange potential.
The B88 exchange potential diverges at the nucleus, whereas the OPM exchange
potential does not diverge.

The radial density $\rho^{\rm av}(r)$ is shown 
in the top part of figure \ref{DFTpot_Mn}.
There are some kinks in the OPM exchange potential. The positions of
the kinks correspond with the positions of troughs in the radial density.
The kinks in the B88 exchange potential are smoother than that in
the OPM exchange potential.
For the X$\alpha$ potential, the kinks are not appeared.

In table \ref{Eng_AOC}, we show the average total energies
(Equation (\ref{average_energy})) using HF method and OPM.
The average total energy of DFT (Equation (\ref{av-Eng-DFT})) 
using X$\alpha$ and Becke's GGA (B88) exchange
functional is also tabulated in table \ref{Eng_AOC}.
The total energy of the OPM method is more closer to the HF total energy
than that of the B88 and X$\alpha$.
The difference between the HF and OPM energies is order of 0.005 au
throughout the transition metals. 

Secondly, we discuss the OPM exchange potentials for 
the 3d$^n$ multiplet states.
The 2-body part of the total energy of an atom (equation (\ref{eng_2body})) 
can be represented as the linear combination of 
$F_\xi^k(ij)$ and $G_\xi^k(ij)$.
For the 3d$^n$ multiplet states,
the coefficients $a_{\rm 3d,3d,2}^\xi$ and $a_{\rm 3d,3d,4}^\xi$ 
depend on the multiplet states, while the other coefficients
do not.
Therefore, the 2-body part of the total energy is expressed as
\begin{eqnarray}
\label{eng_2body_sep}
E_2^\xi = E^\xi_0 
+ a_{\rm 3d,3d,2}^\xi F_\xi^2({\rm 3d,3d})
+ a_{\rm 3d,3d,4}^\xi F_\xi^4({\rm 3d,3d}),
\end{eqnarray}
where $E^\xi_0$ is the 3d$^n$ multiplet independent part of 2-body energy. 

Since $E_2^\xi$, $F_\xi^k(ij)$ and $E^\xi_0$ are functionals of
density $n^\xi(r)=\rho^\xi(r)/4\pi r^2$, we can define the potential 
for $E_2^\xi$, $F_\xi^k(ij)$ and $E^\xi_0$ as
$V^\xi_{E_2}(r) = \frac{\delta E_2^\xi}{\delta n^\xi(r)}$,
$V^\xi_{F^k_{ij}}(r) = \frac{\delta F_\xi^k(ij)}{\delta n^\xi(r)}$
and $V^\xi_{E_0}(r) = \frac{\delta E^\xi_0}{\delta n^\xi(r)}$,
respectively.
In the following, we derive the equation to determine these potentials.
We define ${\cal E}^\xi[P^\xi_{n_i l_i}]$
as a general expression of $E_2^\xi$, $F_\xi^k(ij)$ and $E^\xi_0$.
The corresponding potential is defined as
$V^\xi_{{\cal E}}=\frac{\delta {\cal E}^\xi}{\delta n^\xi(r)}$.
From equation (\ref{P-equation}), 
the single electron orbital $P^\xi_{n_i l_i}$ is considered to be
a functional of $V^\xi_{\rm OPM}$. Furthermore, we regard 
$V^\xi_{\rm OPM}$ as a functional of $n^\xi$. Following chain rule of
functional derivative, we get
\begin{eqnarray}
\label{GOEP-tmp1}
V^\xi_{{\cal E}}(r)&=\int dr' \int d\bm{r''} 
\sum_i \frac{\delta {\cal E}^\xi[P^\xi_{n_i l_i}]}{\delta P^\xi_{n_i l_i}(r')}
\frac{\delta P^\xi_{n_i l_i}(r')}{\delta V^\xi_{\rm OPM}(r'')} 
\frac{\delta V^\xi_{\rm OPM}(r'')}{\delta n^\xi(r)}.
\end{eqnarray}
Multiplying the both sides of equation (\ref{GOEP-tmp1}) by 
$\int d\bm{r} \frac{\delta n^\xi(r)}{\delta V^\xi_{\rm OPM}(r''')}$, and
integrating the right-hand side over $\bm{r}$ and $\bm{r''}$, we obtain
\begin{eqnarray}
\label{GOEP-eq}
\int dr \sum_i 2q_i P^\xi_{n_i l_i}(r)\frac{\delta P^\xi_{n_i l_i}(r)}
{\delta V^\xi_{\rm OPM}(r''')} V^\xi_{{\cal E}}(r)&=
\int dr' \sum_i 
\frac{\delta {\cal E}^\xi[P^\xi_{n_i l_i}]}{\delta P^\xi_{n_i l_i}(r')}
\frac{\delta P^\xi_{n_i l_i}(r')}{\delta V^\xi_{\rm OPM}(r''')},
\end{eqnarray}
where we used the following relation: 
$\int d\bm{r} \frac{\delta n^\xi(r)}{\delta V^\xi_{\rm OPM}(r''')} 
=\int dr \sum_i 2q_i P^\xi_{n_i l_i}(r) \frac{\delta P^\xi_{n_i l_i}(r)}{\delta V^\xi_{\rm OPM}(r''')}$.
If we substitute $E_2^\xi$ for ${\cal E}^\xi[P^\xi_{n_i l_i}]$ in
equation (\ref{GOEP-eq}), we get the OPM equation (\ref{OEPeq-org3}).
Furthermore, if we substitute $F^k_\xi(ij)$ and $E^\xi_0$ for
${\cal E}^\xi[P^\xi_{n_i l_i}]$, equation (\ref{GOEP-eq}) defines 
the potential $V^\xi_{F^k_{ij}}$ and $V^\xi_{E_0}$, respectively.
Since equation (\ref{GOEP-eq}) defines $V^\xi_{{\cal E}}(r)$ only up to a constant,
we set the boundary condition 
$\lim_{r \rightarrow \infty} V^\xi_{\rm OPM}(r) = 0$.
Equation (\ref{GOEP-eq}) is solved using the method described 
in section \ref{cal-method}.

\begin{figure}
\begin{center}
\includegraphics[angle=-90,scale=.6]{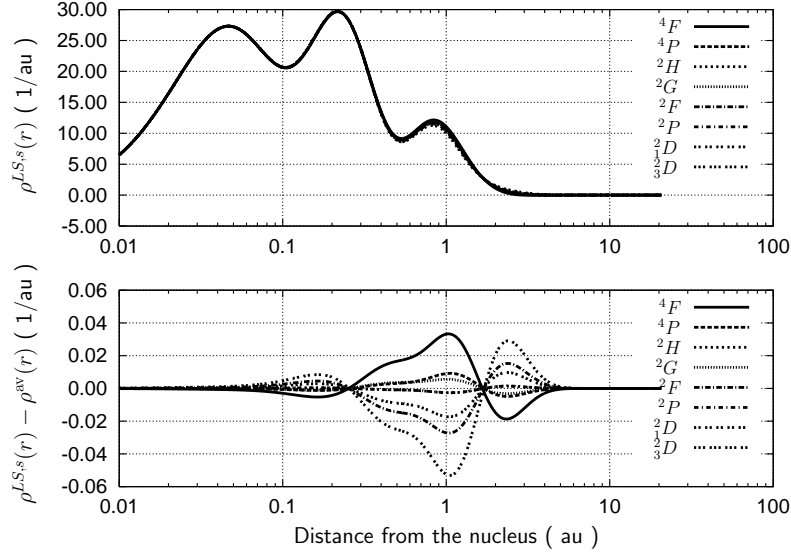}
\caption{
\label{dens_V}
The radial densities $\rho^{LS,s}(r)$ of  V$^{2+}$ multiplet states (top).
The difference between the radial density of the multiplet
 state and that of the average energy: $\rho^{LS,s}(r)-\rho^{\rm av}(r)$ (bottom).
}
\end{center}
\end{figure}

Functional derivative of equation (\ref{eng_2body_sep}) with respect to
$n^{LS,s}(r)$ leads
the decomposition of the OPM potential for
the multiplet state $V^{LS,s}_{\rm OPM}(r)$;
\begin{eqnarray}
V_{\rm OPM}^{LS,s}(r) &= \frac{\delta E_2^{LS,s}}{\delta n^{LS,s}(r)}
= V^{LS,s}_{E_0}(r) 
+a_{\rm 3d,3d,2}^{LS,s} V^{LS,s}_{F^2_{\rm 3d,3d}}(r)
+a_{\rm 3d,3d,4}^{LS,s} V^{LS,s}_{F^4_{\rm 3d,3d}}(r).
\label{sep-V}
\end{eqnarray}
In the top part of figure \ref{dens_V}, we show
the radial density $\rho^{LS,s}(r)$ of each multiplet states of V$^{2+}$.
In the bottom part of figure \ref{dens_V}, we display the
difference of the radial density
$\rho^{LS,s}(r)-\rho^{\rm av}(r)$.
Comparing the top and the bottom parts of figure \ref{dens_V}, 
we find that the difference between the radial densities of 
the multiplet states are much
smaller than the radial densities.
So we neglect the difference of the radial components $P_{n_i l_i}^\xi$ 
between the average configuration and the multiplet states, 
and we substitute
$V^{\rm av}_{F^2_{\rm 3d,3d}}(r)$ and $V^{\rm av}_{F^4_{\rm 3d,3d}}(r)$
for
$V^{LS,s}_{F^2_{\rm 3d,3d}}(r)$ and $V^{LS,s}_{F^4_{\rm 3d,3d}}(r)$ in
equation (\ref{sep-V}), respectively.
The difference between the OPM exchange potential 
for the multiplet states $V_{\rm ex}^{LS,s}(r)$ and 
that for the average energy $V_{\rm ex}^{\rm av}(r)$ can be approximated as
a linear combination of $V_{F^2_{\rm 3d,3d}}^{\rm av}(r)$ and
$V_{F^4_{\rm 3d,3d}}^{\rm av}(r)$:
\begin{eqnarray}
V_{\rm ex}^{LS,s}(r) -V_{\rm ex}^{\rm av}(r) &\simeq&
 (a_{\rm 3d,3d,2}^{LS,s} - a_{\rm 3d,3d,2}^{\rm av}) V^{\rm av}_{F^2_{\rm 3d,3d}}(r)
\nonumber\\
&&+(a_{\rm 3d,3d,4}^{LS,s} - a_{\rm 3d,3d,4}^{\rm av}) V^{\rm av}_{F^4_{\rm 3d,3d}}(r).
\label{ex-lc}
\end{eqnarray}
Neglecting the multiplet dependence of the orbital, we can replace the orbital
$P^\xi_{n_i,l_i}$ in equation (\ref{eng_1body_sub}) and
 (\ref{eng_2body_sep}) to $P^{\rm av}_{n_i,l_i}$. 
Then, the 1-body and 2-body energies are approximated as
\begin{equation}
\begin{array}{l}
\label{eng_2body_sep_av}
E^\xi_1 \simeq E^{\rm av}_1 = \sum_i q_i I^{\rm av}(n_i l_i)\\
E^\xi_2 \simeq E^{\rm av}_0 
+ a_{\rm 3d,3d,2}^\xi F_{\rm av}^2({\rm 3d,3d})
+ a_{\rm 3d,3d,4}^\xi F_{\rm av}^4({\rm 3d,3d}).
\end{array}
\end{equation}

\begin{figure}
\begin{center}
\includegraphics[angle=-90,scale=.6]{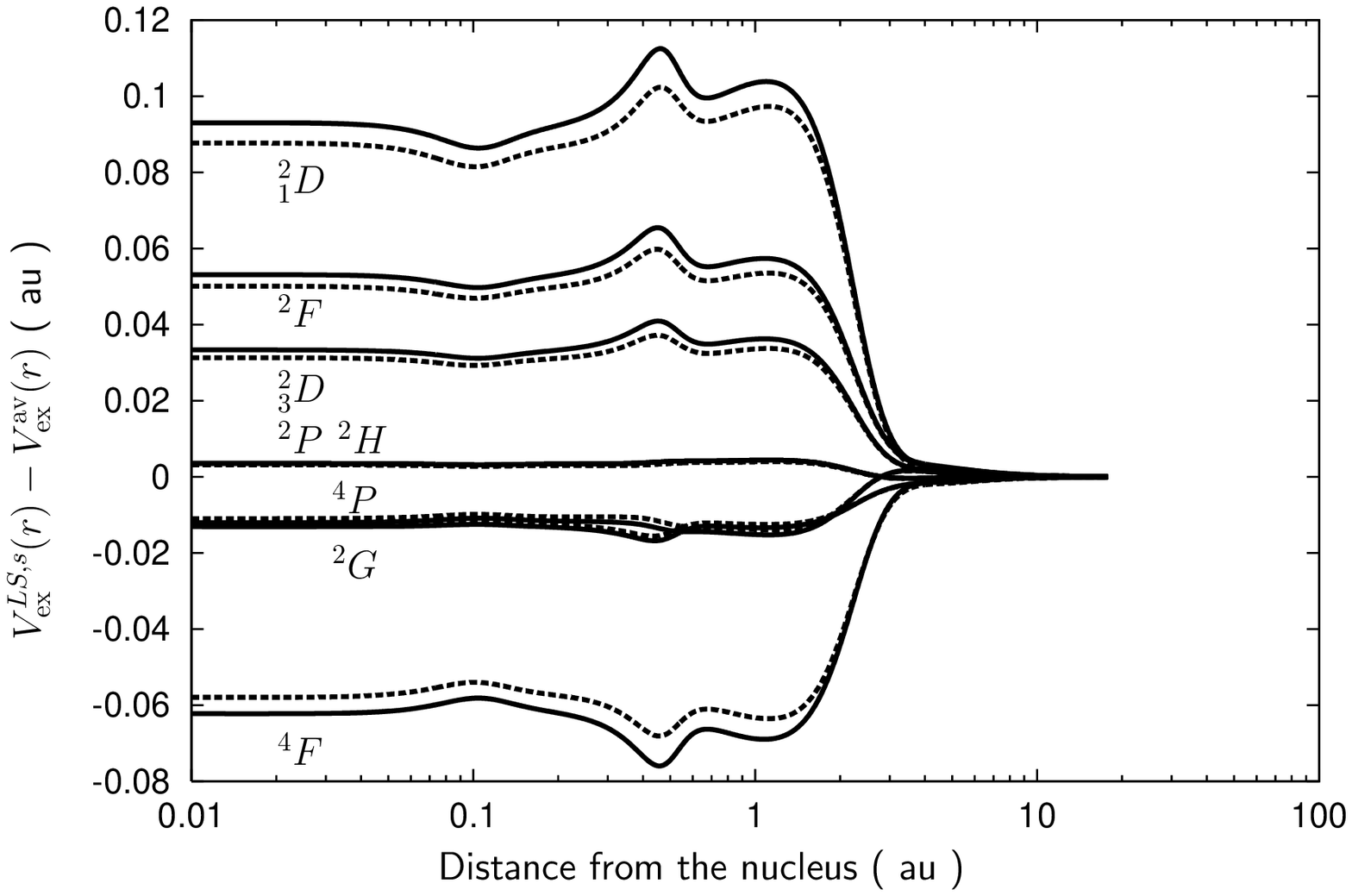}
\caption{
\label{F2F4LC_V}
The difference between the OPM exchange potential of V$^{2+}$ 
multiplet state ($V_{\rm ex}^{LS,s}$)
and that of the average energy ($V_{\rm ex}^{\rm av}$):  $V_{\rm ex}^{LS,s}-V_{\rm ex}^{\rm av}$.
The dotted line is the difference potential 
calculated from equation (\ref{ex-lc}).
The solid line is the same difference calculated from exact equation 
(\ref{sep-V}).
}
\end{center}
\end{figure}

In order to discuss the equality of equations 
(\ref{ex-lc}) and (\ref{eng_2body_sep_av}), we compare the approximate OPM
exchange potentials of V$^{2+}$ multiplets with exact ones.
The dotted line of figure \ref{F2F4LC_V} exhibits the difference between
the OPM exchange potential of the multiplet states and that of the average energy calculated from equation (\ref{ex-lc}).
In the solid line of figure \ref{F2F4LC_V}, we show the same difference
calculated from exact equation (\ref{sep-V}).
The dotted line of figure \ref{F2F4LC_V} close to the solid line.
Small differences come from the neglect of the multiplet
dependence of the radial wave functions.

In third column of table \ref{Eng_Mult}, we show the approximate total
energy of V$^{2+}$ multiplets calculated 
from equation (\ref{eng_2body_sep_av}). For
comparison, the exact OPM total energies are shown in 
second column of table \ref{Eng_Mult}.
If the total multiplet energies are deviate from the average of configuration, 
the approximate equation (\ref{eng_2body_sep_av}) poorly predicts 
the total energies.
%
However, the differences between the approximate and the exact 
total energies are no more than 0.002 au. 
For the other 3d$^n$ transition metal atoms, the differences 
between the approximate and the exact energies are no more than 0.005 au.
Therefore, we can conclude that the equation (\ref{ex-lc}) and
 (\ref{eng_2body_sep_av}) work as the approximation formula 
of the exchange potential and the total energy for the multiplet states,
 respectively.

\begin{figure}
\begin{center}
\includegraphics[angle=-90,scale=.6]{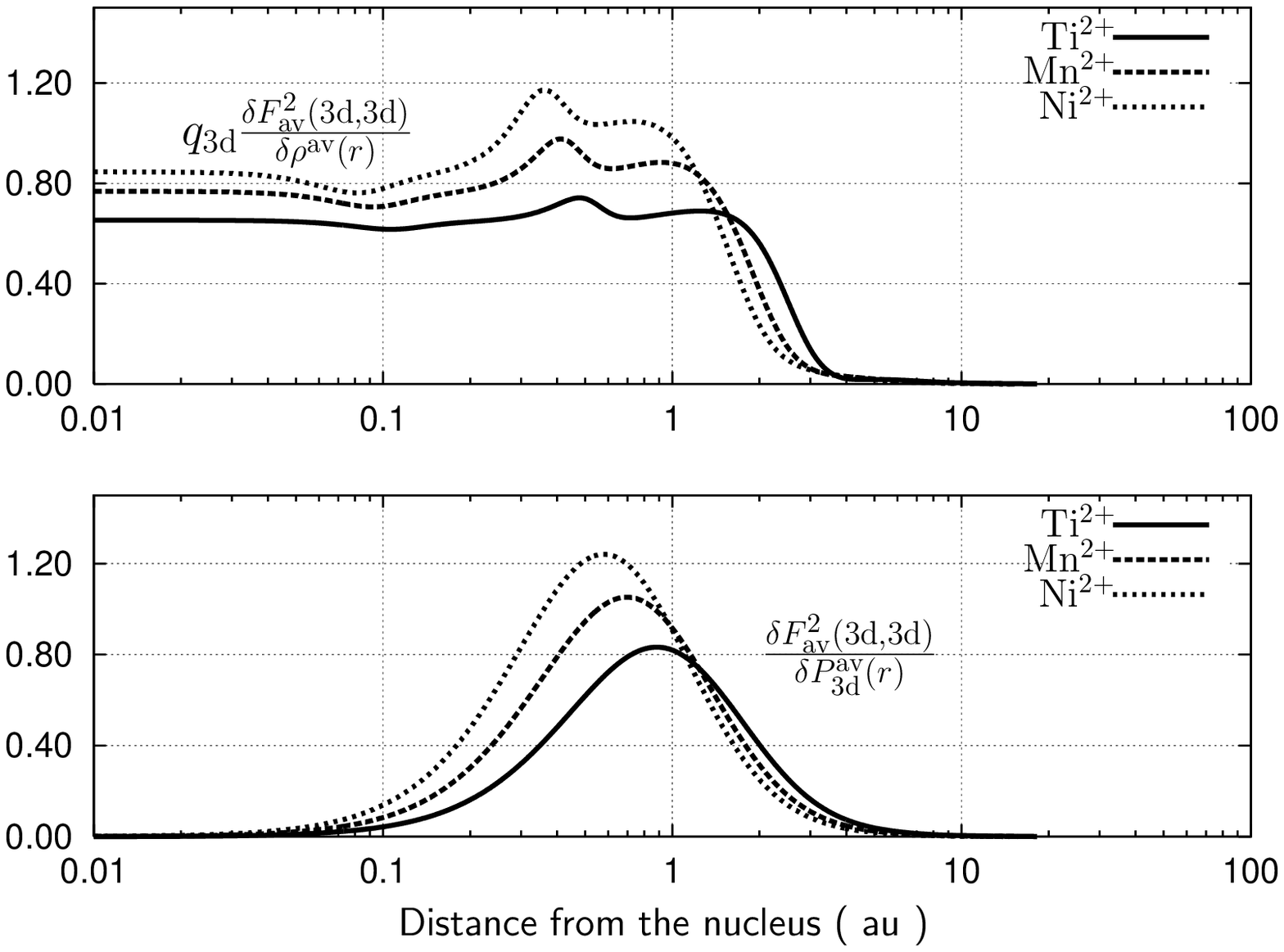}
\caption{
\label{F2pot}
The potential of the Slater integral $F^2_{\rm av}({\rm 3d,3d})$ 
multiplied by the occupation number: $q_{\rm 3d} V^{\rm av}_{F^2_{\rm 3d,3d}}$ (top).
The functional derivative of the Slater integral $F^2_{\rm av}({\rm 3d,3d})$ 
with respect to $P_{\rm 3d}^{\rm av}$: 
$\frac{\delta F^2_{\rm av}({\rm 3d,3d})}{\delta P_{\rm 3d}^{\rm av}}$ (bottom).
}
\end{center}
\end{figure}

\begin{figure}
\begin{center}
\includegraphics[angle=-90,scale=.6]{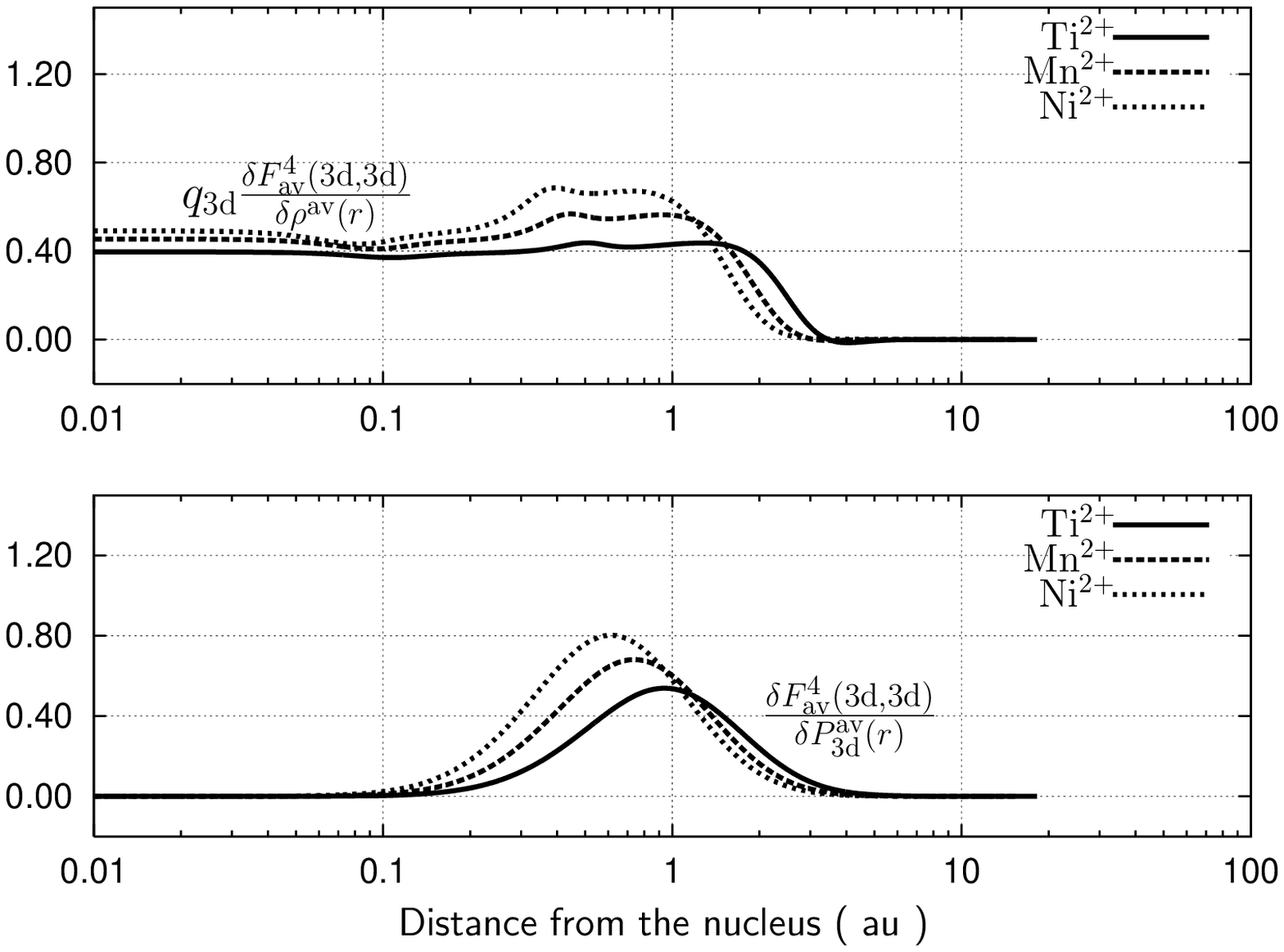}
\caption{
\label{F4pot}
The potential of the Slater integral $F^4_{\rm av}({\rm 3d,3d})$ 
multiplied by the occupation number: $q_{\rm 3d} V^{\rm av}_{F^4_{\rm 3d,3d}}$ (top).
The functional derivative of the Slater integral $F^4_{\rm av}({\rm 3d,3d})$ 
with respect to $P_{\rm 3d}^{\rm av}$: 
$\frac{\delta F^4_{\rm av}({\rm 3d,3d})}{\delta P_{\rm 3d}^{\rm av}}$ (bottom).
}
\end{center}
\end{figure}

In stead of showing the OPM potentials for many multiplet states,
we illustrated $V_{F^2_{\rm 3d,3d}}^{\rm av}(r)$ and 
$V_{F^4_{\rm 3d,3d}}^{\rm av}(r)$ multiplied by
the occupation number $q_{\rm 3d}$ in the top parts of figure \ref{F2pot}
and  figure \ref{F4pot}, respectively.
As atomic number increases, the potentials
$q_{\rm 3d}V_{F^2_{\rm 3d,3d}}^{\rm av}(r)$ and
$q_{\rm 3d}V_{F^4_{\rm 3d,3d}}^{\rm av}(r)$ shrink and become high.
This corresponds with the spread tendency of the electron density or the
single electron orbital.
In the bottom parts of figure \ref{F2pot} and figure \ref{F4pot}, we show  
functional derivative of $F^2_{\rm av}({\rm 3d,3d})$ and 
$F^4_{\rm av}({\rm 3d,3d})$ with respect to
$P_{\rm 3d}^{\rm av}(r)$, respectively.
Maximum of $\frac{\delta F^k_{\rm av}({\rm 3d,3d})}{\delta P^{\rm av}_{\rm 3d}}$
is approximate to that of $q_{\rm 3d}V_{F^2_{{\rm 3d,3d}}}^{\rm av}(r)$, but 
the shapes of the potentials are different.
The two peaks are observed in $q_{\rm 3d}V_{F^2_{\rm 3d,3d}}^{\rm av}(r)$, 
whereas, a single peak in 
$\frac{\delta F^k_{\rm av}({\rm 3d,3d})}{\delta P^{\rm av}_{\rm 3d}}$.
Furthermore, at the nucleus, 
$\frac{\delta F^k_{\rm av}({\rm 3d,3d})}{\delta P^{\rm av}_{\rm 3d}}$ and
$q_{\rm 3d}V_{F^2_{\rm 3d,3d}}^{\rm av}(r)$ are zero and nonzero value, 
respectively.

Finally, we discuss the relation between the OPM exchange potential for the
multiplet state and that for the single determinant.
The multiplet states $\Psi(LSL_z S_z,s)$
are represented as the linear combination of the single determinants
$\Phi(L_z S_z,\nu)$.
If the radial components of the wavefunctions do not depend on 
the multiplet states, 
the single determinant $\Phi(L_z S_z,\nu)$ is represented as
\begin{eqnarray}
\Phi(L_z S_z, \nu) =
\sum_{LS}\sum_s \alpha^{\nu}_{LS L_z S_z,s} \Psi(LSL_z S_z,s).
\end{eqnarray}
The 2-body energy of the single determinant $E_2^{L_z S_z, \nu}$
is the expectation value of $\sum_{i,j}\frac{1}{r_{ij}}$ with respect to
$\Phi(L_z S_z, \nu)$, which can be expressed as
\begin{eqnarray}
\label{e2slmul}
E_2^{L_z S_z, \nu} =
\sum_{LS} \sum_{s,s'} 
\alpha^{\nu*}_{LS L_z S_z,s} \alpha^{\nu}_{LS L_z S_z,s'}
E^{LS,s}_2,
\end{eqnarray}
where $E^{LS,s}_2=\int \Psi^*(LSL_z S_z,s') \sum_{ij}\frac{1}{r_{ij}}
\Psi(LSL_z S_z,s) d\bm{\tau}_1\cdots d\bm{\tau}_N$, and
$\bm{\tau}$ is the spatial and spin coordinates.
Selecting some combination of single determinants, we solve 
simultaneous equations (\ref{e2slmul}) with respect to the 2-body part of 
the multiplet state. The solution of the simultaneous equations 
is approximately valid because the
orbital wave functions depend on multiplet states or single determinants.
\begin{eqnarray}
\label{energy_decomp}
E_2^{LS,s} \simeq \sum_{L_zS_z,\nu} \beta_{L_zS_z,\nu} E_2^{L_z S_z, \nu}.
\end{eqnarray}
Replaceing $\sum_{i,j} \frac{1}{r_{ij}}$ appearing in the derivation of 
(\ref{energy_decomp}) with total Hamiltonian, 
we obtain approximate equation for the total energy.
\begin{eqnarray}
\label{total_energy_decomp}
E^{LS,s} \simeq \sum_{L_zS_z,\nu} \beta_{L_zS_z,\nu} E^{L_z S_z, \nu}.
\end{eqnarray}

\begin{figure}
\begin{center}
\includegraphics[angle=-90,scale=.6]{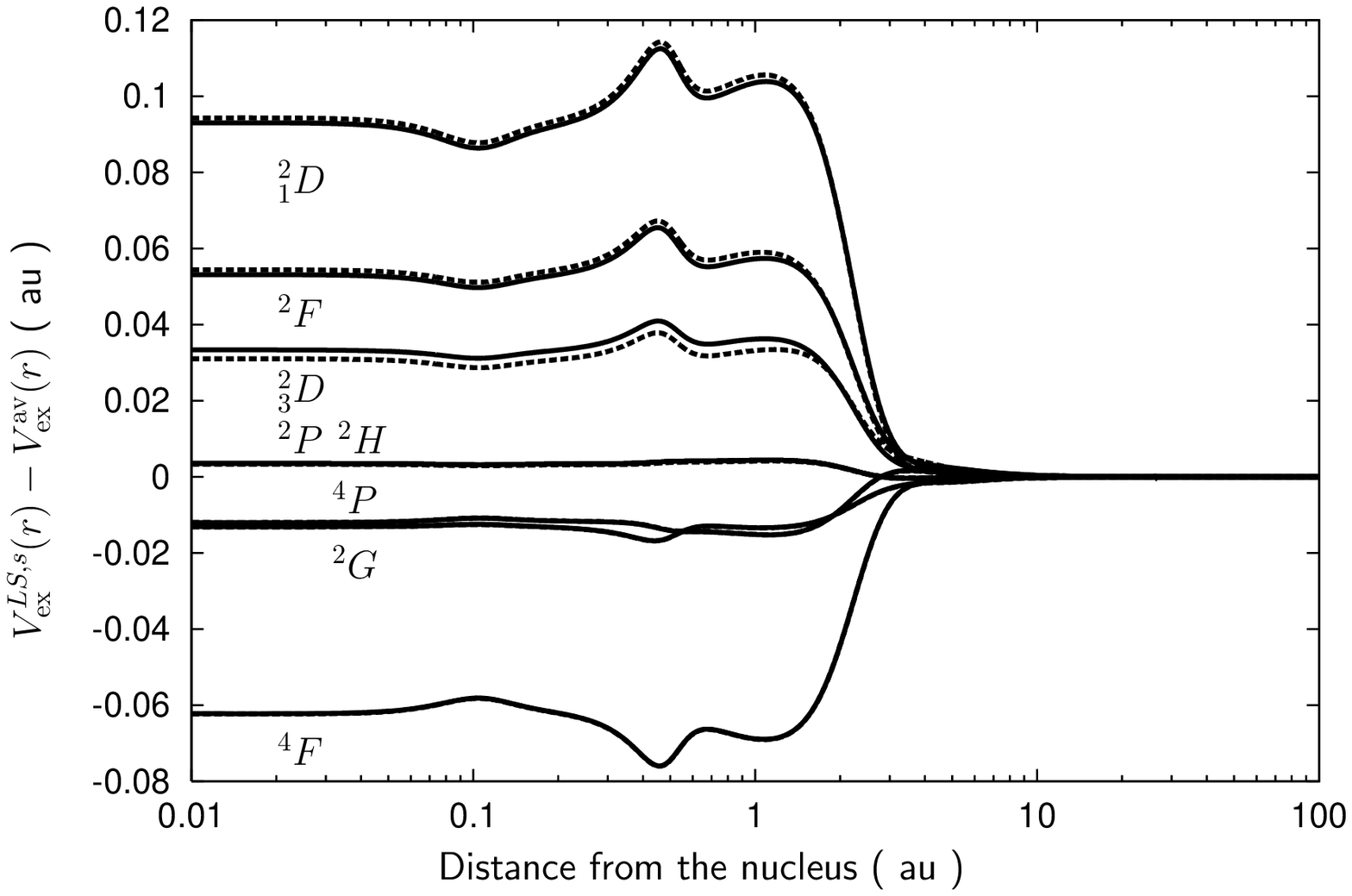}
\caption{
\label{OEPdecomp_V}
The difference between the OPM exchange potential 
of the V$^{2+}$ multiplet state ($V_{\rm ex}^{LS,s}$)
and that of the average energy ($V_{\rm ex}^{\rm av}$):
$V_{\rm ex}^{LS,s}-V_{\rm ex}^{\rm av}$.
The dotted line is the difference potential
 calculated 
from equation (\ref{pot_decomp}).
The solid line is the same difference 
 calculated 
from exact equation (\ref{sep-V}).
}
\end{center}
\end{figure}

From the variation of equation (\ref{energy_decomp}) with respect to 
$n^{LS,s}(r)$,
the OPM potential for the multiplet state 
$V^{LS,s}_{\rm OPM}(r)=\frac{\delta E_2^{LS,s}}
{\delta n^{LS,s}(r)}$
is expressed as the linear combination of the OPM potentials for the single
determinants $V^{L_z S_z,\nu}_{\rm OPM}(r)=
\frac{\delta E_2^{L_z S_z, \nu}}{\delta n^{LS,s}(r)}$:
\begin{eqnarray}
\label{pot_decomp}
V_{\rm OPM}^{LS,s}(r) \simeq \sum_{L_zS_z,\nu} \beta_{L_zS_z,\nu} 
V_{\rm OPM}^{L_z S_z, \nu}(r).
\end{eqnarray}
Following equation (\ref{pot_decomp}),
we evaluate the approximate OPM exchange potential for the multiplet state.
Figure \ref{OEPdecomp_V} presents the difference between the OPM potential
for V$^{2+}$ multiplet state and that for the average energy. The dotted line
exhibits the approximate difference potential calculated 
from equation (\ref{pot_decomp}). 
We select the single determinant appearing in equation (\ref{pot_decomp}) 
as 
$|2,1,-2|$, $|2,1,\bar{1}|$, $|2,1,\bar{0}|$, $|2,1,\bar{-1}|$, 
$|2,-2,\bar{2}|$, $|1,0,\bar{1}|$, $|2,1,\bar{-2}|$, $|2,1,0|$, 
$|2,1,\bar{2}|$, 
where the number represents the $z$ component of orbital angular momentum.
The absence and presence of the line over the number stand for spin up
and down state, respectively.
In the solid line of figure \ref{OEPdecomp_V}, 
we show the difference potential calculated from exact equation (\ref{sep-V}).
%
%
We find, from figure \ref{OEPdecomp_V}, 
that the approximate difference potential ( dotted line ) 
is very close to the exact one ( solid line ).

In the fourth column of table \ref{Eng_Mult}, we show 
the approximate total energy calculated from equation
(\ref{total_energy_decomp}). 
We compare the approximate total energy with exact 
 OPM total energy presented in second column of 
table \ref{Eng_Mult}.
%
The difference between the approximate total energy 
and the exact one is no more than 0.004 au. 
For the other 3d transition metal atoms, the difference between two
energies is no more than 0.01 au.
Therefore, the exchange potential for the multiplet state can be
evaluated using the theory such as the density functional theory, 
where the electronic structure is based on the single determinant,

\section{Conclusion}

We applied OPM to the multiplet energies of 
the 3d transition metal atoms using the total energy functional of the
single-configuration HF \cite{Fischer} form.
The calculated OPM exchange potential can be approximated by the
$X\alpha$ and B88 exchange potentials in $r>0.1$ au.
For near nucleus region ($r<0.1$ au ), the OPM exchange potential 
strongly deviates from the $X\alpha$ and B88 exchange potentials.
The difference between the OPM exchange potential for the multiplet states 
and that for the average energy
is quite smaller than the OPM exchange potential.
The OPM exchange potential for the multiplet state
 of the 3d transition metal atom
can be represented as the linear combination of 
$V^\xi_{F^2_{\rm 3d,3d}},V^\xi_{F^4_{\rm 3d,3d}}$ and $V^\xi_{E_0}$.
We find that the OPM exchange potential can be approximated by
the linear combination of multiplet independent potentials 
$V^{\rm av}_{F^2_{\rm 3d,3d}}$ and $V^{\rm av}_{F^4_{\rm 3d,3d}}$.
This result indicates that the total energy functional 
can be represented as linear combination of the terms
which are the product of the multiplet independent quantities
derived from the Slater integrals 
($F^2_{\rm av}({\rm 3d,3d})$ and $F^4_{\rm av}({\rm 3d,3d})$)
and the multiplet dependent quantities 
($a_{\rm 3d,3d,2}^{LS,s}$, $a_{\rm 3d,3d,4}^{LS,s}$).
%
%

As the representative of the potentials for the many multiplet states 
of the 3d transition metal atoms, we discussed the features of the potentials
$V^{\rm av}_{F^2_{\rm 3d,3d}}$ and $V^{\rm av}_{F^4_{\rm 3d,3d}}$.
We find that the potentials
$V^{\rm av}_{F^2_{\rm 3d,3d}}$ and $V^{\rm av}_{F^4_{\rm 3d,3d}}$ 
shrink, and become high as the atomic number increases, which is
the same trend as observed in 
the wave function of the 3d transition metals.

The multiplet state of the 3d transition metal atom is represented 
as the linear combination of the single determinants.
From our calculation result, we find that the OPM exchange potential
of the multiplet state can be approximated as the linear combination of the OPM
exchange potentials of the single determinants.
The result might be similar to the methods of 
Ziegler \cite{Ziegler}, Wood \cite{Wood} and von Barth \cite{Barth}.
However our result shows that the d electron multiplet state energy
should be evaluated by including multiplet dependency to the total energy
functional.

\ack
One of the authors (N.H.) acknowledges a financial support from 
a Grant from the Ministry of Education, 
Science, Sports and Culture to promote advanced scientific research.

\begin{table}
\caption{\label{Eng_AOC}
The average energy of the configuration (in au ) for 3d$^n$ multiplets
 calculated by Hartree-Fock ($E_{\rm HF}^{\rm av}$), 
OPM ($E_{\rm OPM}^{\rm av}$), Becke 88 ($E_{\rm B88}^{\rm av}$), 
and X$\alpha$ ($E_{\rm X\alpha}^{\rm av}$,$\alpha=2/3$) 
method.}
\begin{tabular}{rrrrr}
\hline
& $E_{\rm HF}^{\rm av}$ & $E_{\rm OEP}^{\rm av}-E_{\rm HF}^{\rm av}$ 
& $E_{\rm B88}^{\rm av}-E_{\rm HF}^{\rm av}$ 
& $E_{\rm X\alpha}^{\rm av}-E_{\rm HF}^{\rm av}$ \\
\hline
Ti$^{2+}$ & -847.6927 & 0.0061 &  0.0096 & 2.9047 \\
 V$^{2+}$ & -942.0952 & 0.0063 & -0.0059 & 3.0478 \\
Cr$^{2+}$ &-1042.4387 & 0.0065 & -0.0257 & 3.1876 \\
Mn$^{2+}$ &-1148.8609 & 0.0066 & -0.0492 & 3.3252 \\
Fe$^{2+}$ &-1261.4995 & 0.0067 & -0.0754 & 3.4613 \\
Co$^{2+}$ &-1380.4916 & 0.0068 & -0.1034 & 3.5968 \\
Ni$^{2+}$ &-1505.9743 & 0.0068 & -0.1320 & 3.7330 \\
\hline
\end{tabular}
\end{table}

\begin{table}
\caption{\label{Eng_Mult} The total energies (in au ) of V$^{2+}$ 
multiplet states calculated by the HF
 method, OPM, equation (\ref{eng_2body_sep_av}) and
 equation (\ref{total_energy_decomp}).}
\begin{tabular}{lllll}
\hline
Multiplet & HF & OPM & Eq. (\ref{eng_2body_sep_av}) & Eq. (\ref{total_energy_decomp})\\
\hline
 ${}^4$F &    -942.1799 &    -942.1733 & -942.1727 & -942.1733\\
 ${}^4$P &    -942.1128 &    -942.1065 & -942.1065 & -942.1064\\
 ${}^2$H &    -942.0902 &    -942.0840 & -942.0840 & -942.0840\\
 ${}^2$G &    -942.1124 &    -942.1061 & -942.1060 & -942.1060\\
 ${}^2$F &    -942.0245 &    -942.0181 & -942.0177 & -942.0161\\
 ${}^2$P &    -942.0902 &    -942.0840 & -942.0840 & -942.0842\\
 ${}^2_1$D &    -941.9693 &    -941.9628 & -941.9612 & -941.9606\\
 ${}^2_3$D &    -942.0507 &    -942.0444 & -942.0442 & -942.0478\\
\hline
\end{tabular}
\end{table}

\end{document}